\documentclass{sf2a-conf2015}
\usepackage{graphicx}
\usepackage{hyperref}
\usepackage[]{natbib}  
\usepackage{epstopdf}

\def\BibTeX{{\rm B\kern-.05em{\sc i\kern-.025em b}\kern-.08em
    T\kern-.1667em\lower.7ex\hbox{E}\kern-.125emX}}
\bibpunct{(}{)}{;}{a}{}{,}  

\begin{document}

\TitreGlobal{SF2A 2015}


\title{The optical polarization signatures of fragmented equatorial dusty structures in Active Galactic Nuclei}

\runningtitle{Polarization of clumpy circumnuclear dusty structures}

\author{F. Marin}\address{Astronomical Institute of the Academy of Sciences, Bo{\v c}n\'{\i} II 1401, CZ-14100 Prague, Czech Republic}

\author{M. Stalevski$^{2,3,}$}\address{Departmento de Astronomia, Universidad de Chile, Camino del Observatorio 1515, Santiago, Chile}
\address{Astronomical Observatory, Volgina 7, 11060 Belgrade, Serbia}\address{Sterrenkundig Observatorium, Universiteit Gent, Krijgslaan 281-S9, Gent, 9000, Belgium}

\setcounter{page}{237}


\maketitle

\begin{abstract}
If the existence of an obscuring circumnuclear region around the innermost regions of active galactic nuclei 
(AGN) has been observationally proven, its geometry remains highly uncertain. The morphology usually adopted 
for this region is a toroidal structure, but other alternatives, such as flared disks, can be a good representative 
of equatorial outflows. Those two geometries usually provide very similar spectroscopic signatures, even when
they are modeled under the assumption of fragmentation. In this lecture note, we show that the resulting 
polarization signatures of the two models, either a torus or a flared disk, are quite different from each 
other. We use a radiative transfer code that computes the 2000 -- 8000~\AA~polarization of the two morphologies 
in a clumpy environment, and show that varying the sizes of a toroidal region has deep impacts onto the 
resulting polarization, while the polarization of flared disks is independent of the outer radius. Clumpy 
flared disks also produce higher polarization degrees ($\sim$~10~\% at best) together with highly variable 
polarization position angles.
\end{abstract}



\section{Introduction}
The presence of an optically thick, equatorial, dusty region around the core of active galactic nuclei (AGN) has been 
predicted by polarimetric observations from the Lick 3m telescope \citep{Antonucci1985} and revealed thanks 
to mid-infrared interferometry \citep{Jaffe2004,Wittkowski2004} at the \textit{Very Large Telescope}. This obscuring
torus is a fundamental region that explains the polarization dichotomy observed in radio-quiet AGN: along type-1
viewing angles (close to the AGN pole), the observed net polarization position angle is mainly parallel to 
the symmetry axis of the system, while at type-2 orientations (close to the edge of the system), the polarization angle 
is perpendicular. Hiding the innermost regions of AGN behind a dust wall is also necessary to explain the absence of 
broad lines in the optical spectra of type-2 Seyfert galaxies. It has been shown by \citet{Antonucci1985} and 
\citet{Miller1990} that polarization can unveil those broad lines in the polarized flux spectra of type-2 AGN, 
demonstrating the power of polarimetry.

Studying AGN under the scope of polarization can be achieved thanks to numerical modeling (e.g. 
\citealt{Kartje1995,Wolf1999,Young2000}), but all those codes are limited to scattering inside uniform-density regions. 
As a bulky hydrostatic dusty region is inconsistent with self-gravitational stability, it is more likely to consist 
of individual, optically thick, molecular clumps in collision-free orbits that are sustaining the vertical height of 
the torus \citep{Krolik1988,Pier1992}. Hence, it is necessary to use radiative transfer codes that are able to handle 
multiple scattering in a clumpy environment, such as {\sc skirt} in the optical and infrared bands \citep{Baes2011,Stalevski2012,Camps2015}, 
or {\sc stokes} in the optical, ultraviolet and X-ray bands \citep{Goosmann2007,Marin2012}. The later code has been 
recently upgraded to account for thousands of individual scattering regions \citep{Marin2015} and will be used here 
to investigate the impact of fragmentation onto two different geometries of circumnuclear regions.

In this lecture note, the optical polarimetric investigation of dusty clumpy tori and flared disks will be presented
in Sect.~\ref{Tori} and Sect.~\ref{Disks}, respectively. We discuss our results and conclude this note in 
Sect.~\ref{Conclusion}.

\section{Polarization modeling}
\label{Introduction}
We used the Monte Carlo radiative transfer code {\sc stokes} \citep{Goosmann2007,Marin2012,Marin2015} to compute the 
optical polarization of two different equatorial models, using torus and flared disk morphologies. Both geometries
are investigated using a clumpy medium composed of hundreds of constant-density spheres filled with a Milky Way dust 
mixture. Each optically-thick clump has a radius of 0.3~pc and an optical depth in the V-band of 50. We fixed 
the filling factor of the models to 25~\%. Both geometries have an inner radius of 0.25~pc, a distance to the 
accretion disk corresponding to the dust sublimation radius for L$_{\rm AGN} \approx$ 4 $\times$ 10$^{44}$~erg~s$^{-1}$, 
assuming T = 1500~K for the dust sublimation temperature \citep{Barvainis1987}. Note that, since here we are interested in 
polarization only, and not thermal dust re-emission, the actual AGN luminosity used in the models does not matter.
The outer radius of the torus is not fixed by the unified model of AGN and is thought to span from a couple of parsecs 
to about 100~pc. Recent observations \citep{Gandhi2015} and studies \citep{Marin2015} tend to rule out extended dusty 
structures along the equator but this needs to be investigated further. This is particularly true as infrared spectral 
energy distributions are often not able to provide constraints on the outer dust radius (e.g. \citealt{Alonso2011}). Hence, 
we adopt two different outer radii for our models; either 100~pc or 6~pc \citep{Heymann2012}. We allow the torus half-opening 
angle to vary between 30$^\circ$, 45$^\circ$, and 60$^\circ$, with respect to the symmetry axis of the model. Finally, we 
simulate the irradiating continuum using an isotropic point-like source emitting an unpolarized flux according to power-law 
intensity spectrum $F_{\rm *}~\propto~\nu^{-\alpha}$ with $\alpha = 1$.

\subsection{Fragmented tori}
\label{Tori}
\begin{figure}[t!]
  \centering
  \includegraphics[trim = 10mm 15mm 10mm 20mm, clip, width=0.8\textwidth,clip]{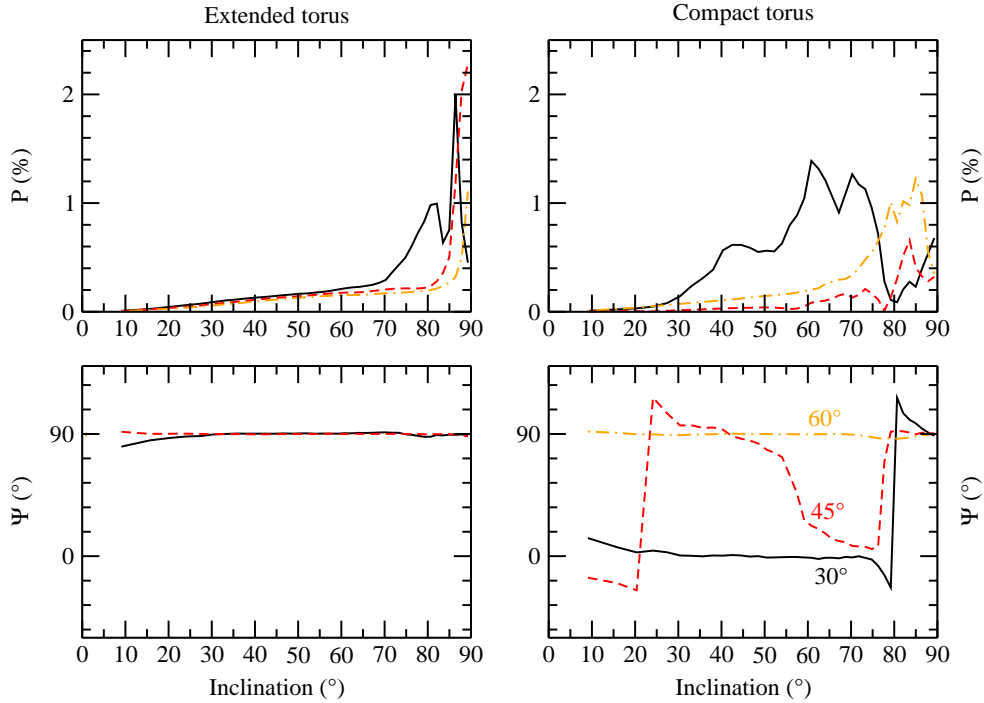}      
  \caption{Polarization degree $P$ (top) and polarization position 
	   angle $\Psi$ (bottom) for an extended (left column) 
	   and a compact clumpy torus (right column). 
	   Polarization is plotted as a function of the observer's 
	   viewing angle for three different half-opening angles
	   of the torus (solid black line: 30$^\circ$, dashed 
	   red line: 45$^\circ$, and dot-dashed orange line: 60$^\circ$), 
	   defined with respect to the symmetry axis of the model.}
  \label{Fig:Torus}
\end{figure}

The resulting 2000 -- 8000~\AA~polarization degree $P$ and polarization angle $\Psi$ of different fragmented tori 
are integrated and plotted in Fig.~\ref{Fig:Torus} against the inclination of the observer. An inclination of 
0$^\circ$ corresponds to a pole-on view and 90$^\circ$ corresponds to an edge-on viewing angle.

In the case of an extended clumpy torus (Fig.~\ref{Fig:Torus} left column), the resulting polarization $P$ increases 
with inclination, up to a maximum value that depends on the half-opening angle of the system. A maximum of 2~\% is 
detected at extreme inclinations for tori with large half-opening angles. This is the result of multiple scattering 
between the dust clouds in an environment that enables radiation to escape even at inclinations where obscuration 
is maximum. The associated polarization position angle is equal to 90$^\circ$, i.e. parallel to the projected 
symmetry axis of the torus. This is a geometrical effect predicted by \citet{Kartje1995} and already reproduced 
in \citet{Marin2015}, where photons scatter off the side walls of the dust structure and naturally yield parallel
$\Psi$ values.

Looking at compact dusty tori (Fig.~\ref{Fig:Torus} right column), the maximum value of $P$ is lower ($\sim$~1~\%)
but does not peak as in the case of extended tori. This is due to the compactness of the model: the torus being 
smaller in diameter while sustaining a similar height (scaled with the half-opening angle), the region has a less 
oblate morphology. Hence, the angular dependence of opacity provides higher obscuration at intermediate inclinations 
and radiation escapes with more difficulty from the funnel. This geometrical effect also has an impact on the 
polarization angle as photons do not longer scatter preferentially along the midplane, resulting in polarization 
position angles orthogonal to the scattering plane for the cases where the half-opening angle of the model is lower 
than 60$^\circ$. At large opening angles, where the resulting structure is geometrical flat, scattering will naturally 
produce parallel polarization angles.

\subsection{Fragmented flared disks}
\label{Disks}

The geometry of a flared disk is completely different from a torus, as the ratio of the disk thickness to the distance 
from the central black hole increases outward. Hence, we expect different signatures in terms of polarization.

\begin{figure}[t!]
  \centering
  \includegraphics[trim = 10mm 15mm 10mm 20mm, clip, width=0.8\textwidth,clip]{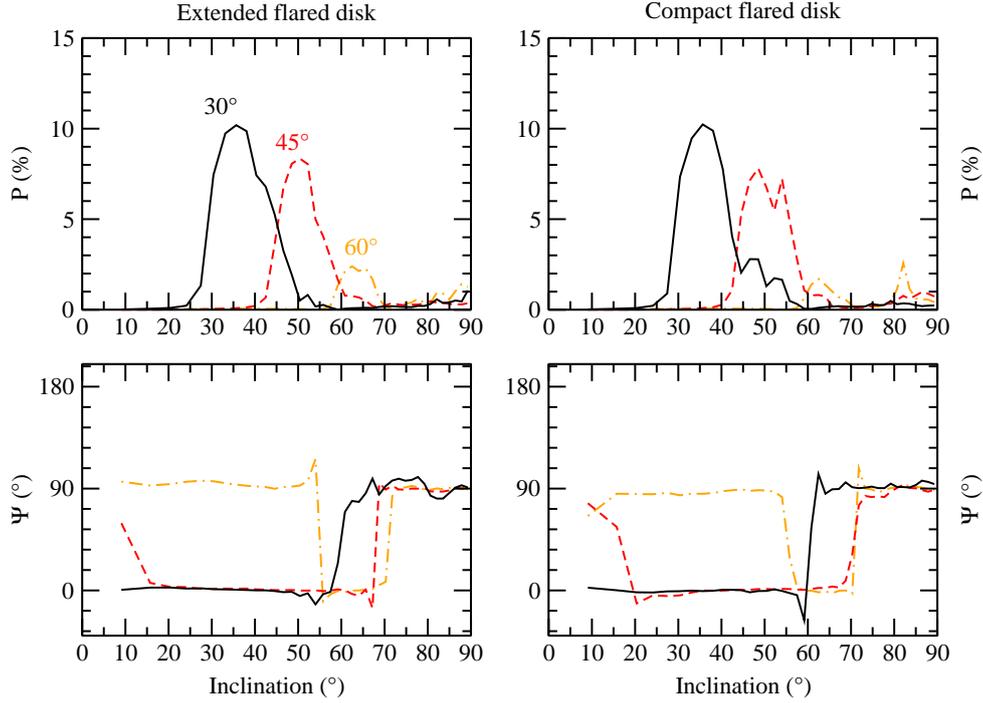}      
  \caption{Same as Fig.~\ref{Fig:Torus}, but for fragmented
	   flared disks.}
  \label{Fig:Flared}
\end{figure}

Fig~\ref{Fig:Flared} (left column) shows the polarization resulting from an extended flared disk. $P$ drastically increases 
at viewing angles that matches the half-opening angle of the flared disk, with a maximum polarization degree of $\sim$~10~\% 
for a geometrically puffed-up region. The low $P$ before this peculiar inclination is due to a direct view of the central 
engine, that emits unpolarized photons, hence diluting the net polarization percentage. The decrease of $P$ when the observer's 
line-of-sight exceeds the flared disk horizon is due to canceling contribution of radiation with a perpendicular polarization angle 
scattered from the inner walls of the disk and photons scattered along the midplane, thus carrying a parallel $\Psi$.
This trend can be observed in the polarization angle plot (Fig~\ref{Fig:Flared} bottom left), where the competition between 
parallel and perpendicular polarization drives the resulting $P$. At maximum inclinations, $\Psi$ = 90$^\circ$, such as 
in the torus cases, for the same reasons: the gaps between the clouds allow photons that have scattered close to the equator 
to escape, carrying a parallel polarization position angle.

Focusing on compact flared disks (Fig~\ref{Fig:Flared} right column), we find very similar results. This is due to the fact that
reducing the outer radius of the dusty region does not change its global geometry. The amount of $P$ and the inclination-dependent 
behavior of $\Psi$ are thus the same as in the case of an extended flared disk.

\section{Discussion and conclusion}
\label{Conclusion}
The radiative transfer modeling undertaken in this lecture note confirms that polarization is a unique tool to distinguish between 
various dusty equatorial morphologies. The polarization signatures of clumpy tori and flared disks qualitatively and quantitatively
show contrasts: a torus cannot produce as high degrees of polarization as a flared disk (especially for half-opening angles lower 
than 60$^\circ$), and their inclination-dependent $\Psi$ signatures are clearly different. Polarimetry also allows to put constraints 
on the outer radius of toroidal structures as varying their maximal extension impacts their net polarization. In the case of fragmented 
flared disks, changing the outer dust radius results in no polarimetric differences due to the geometry of the system: squeezing its 
width does not alter the maximal height of the disk, in contrast to tori.

The high polarization degrees found for flared disk models (up to 10~\%), when the observer's line-of-sight is grazing the edge of the 
disk, is a feature that might have an impact onto a more complete AGN model. In particular, this could be a hint to explain the high 
($>$ 2~\%) polarization degrees of some type-1 AGN (e.g. Fairall~51, IC~4329A or Mrk~1239, see \citealt{Marin2014}). As shown on 
Fig.~\ref{Fig:Flared} (top row), this would be also consistent with the perpendicular polarization position angle found for most 
(but not all) of those highly polarized Seyfert-1s. It implies that those objects would be seen at a very limited inclination range.
However, this hypothesis must be explored in details with more complete AGN modeling, as the polarimetric results might change when 
equatorial and polar electron scattering regions are added.

\begin{acknowledgements}
This work has been supported by the COST Action MP1104 ``Polarization as a tool to study the Solar System and beyond''. MS acknowledges 
support by FONDECYT through grant No. 3140518 and Ministry of Education, Science and Technological Development of the Republic of Serbia 
through the projects Astrophysical Spectroscopy of Extragalactic Objects (176001) and Gravitation and the Large Scale Structure of the 
Universe (176003).
\end{acknowledgements}

\bibliographystyle{aa}  
\bibliography{marin} 

\end{document}